\documentclass[10pt, conference]{IEEEtran}

\usepackage[colorlinks,urlcolor=blue,linkcolor=blue,citecolor=blue]{hyperref}
\usepackage{color,array}
\usepackage{pdfpages}
\usepackage{graphicx}
\usepackage{epstopdf}
\usepackage{dblfloatfix}
\usepackage{amsmath}
\usepackage{cite}
\usepackage{float}
\usepackage[noend]{algorithm}
\usepackage{algorithmicx}
\usepackage{algpseudocode}

\begin{document}

\title{Edge AI as a Service with Coordinated Deep Neural Networks}

\author{Alireza Maleki, Hamed Shah-Mansouri, and Babak H. Khalaj \\
Department of Electrical Engineering, Sharif University of Technology, Tehran, Iran \\
Email: \{alireza.maleki96, hamedsh, khalaj\}@sharif.edu

\thanks{A. Maleki and H. Shah-mansouri and B. H. Khalaj are with Department of Electrical Engineering, Sharif University of Technology, Tehran, Iran.
Email: {alireza.maleki96, hamedsh, khalaj}@sharif.edu.
}}


\maketitle

\begin{abstract}
As artificial intelligence (AI) applications continue to expand in next-generation networks, there is a growing need for deep neural network (DNN) models. Although DNN models deployed at the edge are promising for providing AI as a service with low latency, their cooperation is yet to be explored. In this paper, we consider that DNN service providers share their computing resources as well as their models' parameters and allow other DNNs to offload their computations without mirroring. We propose a novel algorithm called coordinated DNNs on edge (\textbf{CoDE}) that facilitates coordination among DNN services by establishing new inference paths. CoDE aims to find the optimal path, which is the path with the highest possible reward, by creating multi-task DNNs from individual models. The reward reflects the inference throughput and model accuracy. With CoDE, DNN models can make new paths for inference by using their own or other models' parameters. We then evaluate the performance of CoDE through numerical experiments. The results demonstrate a $40\%$ increase in the inference throughput while degrading the average accuracy by only $2.3\%$. Experiments show that CoDE enhances the inference throughput and, achieves higher precision compared to a state-of-the-art existing method.
\end{abstract}

\vspace{-.2cm}

\section{Introduction}

\begin{figure}[t]
\centering
\includegraphics[width=.4\textwidth,keepaspectratio]{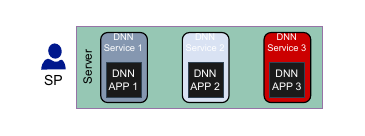}
\label{fig:fig1}
\vspace{-.4cm}
\caption{A SP provides one or multiple DNN services on its server, where each service offers one DNN application (i.e., model).}
\vspace{-.2cm}
\end{figure}

Artificial intelligence (AI) is transforming next-generation networks by offering advanced data analytics and intelligent decision-making capabilities. Network operators can expand the capabilities of resource-limited devices by providing AI services, commonly known as AI as a Service (AIaaS). With the rapid growth of deep neural network (DNN) applications in the AI era, delivering DNN models as services has become even more essential. Fig. \hyperref[fig:fig1]{1} illustrates a service provider (SP) which is hosting three DNN services in its server. DNNs power everything from video analysis and chatbots to autonomous vehicles, gaming \cite{ai-game}, and metaverse \cite{ai-meta}.

\begin{figure*}[h]
\centering
\includegraphics[width=.8\textwidth,keepaspectratio]{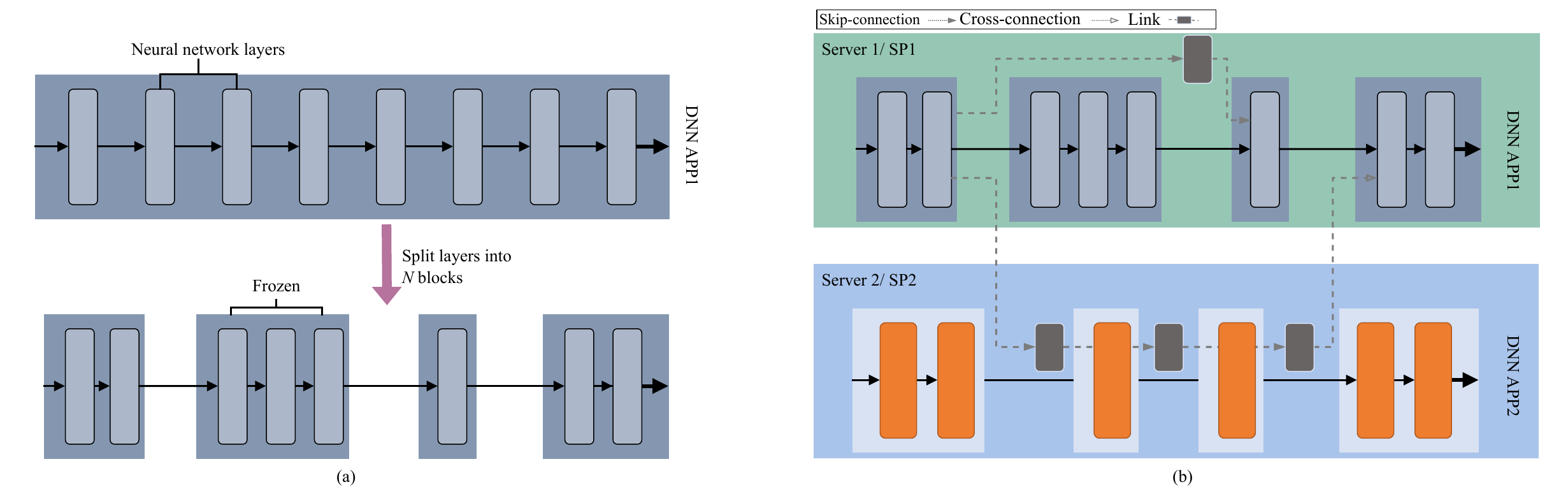}
\label{fig:fig2}
\vspace{-.5cm}
\caption{(a) We consider that SP1 provides its DNN model (i.e., APP1) on server 1. We divide it into a number of manageable blocks. By freezing the model's parameters, SP1 can keep its model integrity through any further training.
(b) In this scenario, SP1 and SP2 provide their services on server1 and server2, respectively. SP1 aims to offload its tasks to server2, and SP2 generates the relative links (i.e., small NN modules) between its blocks. SP1 does not add any links unless it uses skip-connections.}
\vspace{-.5cm}
\end{figure*}

The increasing complexity and computation demands of DNNs lead to a rising need for additional computing resources. Cloud computing platforms with their powerful computing resources \cite{edge-survey-1}, edge computing servers located closer to user devices \cite{edge-early,edge-early-part,CoEdge}, and hybrid cloud-edge environments \cite{spinn,hyb-auto-spl} can provide the high-performance infrastructure required for intensive DNN computation.
Nevertheless, these computing infrastructures are not capable of meeting the ever-growing users' demands. Thus, it has become inevitable to enhance the capabilities of the DNN services. To scale up DNN services on both cloud and edge computing architectures, common techniques such as model distribution, horizontal scaling, and replication are used \cite{CoEdge},\cite{hyb-auto-spl,edge-micro}. Existing methods isolate DNN services, requiring dedicated resources and pipelines. This isolation also necessitates a complete training if services want to share a model.  However, some SPs might have correlated services that could share models and resources, enhancing capabilities without significant resource increase. In other words, a DNN service, which we call the \textit{host} service, can use its model parameters and computing resources to perform a part of the task of another DNN service, referred to as the \textit{local} service,  as shown in Fig. \hyperref[fig:fig2]{2}. This sharing can occur within an SP or across different providers. SPs can create extra inference paths for their DNN services by using other DNN services' models while introducing only a small number of learning parameters.
A path is a sequence of neural network blocks, where the blocks construct the DNN models.

\vspace{-.05cm}
\subsection{Related Work}
Multi-task learning (MTL) is a technique in which a shared model is used to learn multiple tasks simultaneously \cite{multi-survey}. MTL leverages task correlations to efficiently share representations and process high-level features among multiple tasks. Neural architecture search methods explore architectural space to optimize MTL model performance, incorporating branches, skip-connections \cite{switch}, model refinements, or building a model from scratch \cite{mtl-search}. Researchers have developed various techniques for constructing MTL models. In \cite{mtl-nas}, a feature fusion technique is used to combine two distinct models to improve the overall performance. BERT models \cite{bert} use adapter modules \cite{adapter} to adapt pre-trained knowledge to new tasks and enable task-specificity without modifying the base models' parameters. We use this concept to change data representation when passing data to other inference paths.

Another technique to reduce the computational complexity of DNN models and improve their efficiencies is the early exits (EE) method. With this approach, a number of exit branches are added among the model layers. These EE branches make alternative pathways for the DNN services and can reduce the computation for each task with the cost of reduced precision and increased model parameters. Data is routed through the exit branches when the application is time-sensitive or the system is under heavy load \cite{edge-early-part}. Alongside EE branches, current methods like SPINN \cite{spinn} employ synergistic approaches to distribute computation among end devices, edge nodes, and cloud services. It requires the sharing of DNN models and increases resource usage for service replication, posing challenges when computing resources are exhausted.

Instead of offloading entire tasks, we can split models across several devices. Each device works on a part of the model and then sends its results to a central unit that combines everything for the final inference. In \cite{Aggregation}, data is sent across devices with smaller models, halting computation upon acceptable certainty or forwarding outputs to edge nodes. Similarly, in \cite{CoEdge}, models are distributed across devices, each performing a portion of computation and forwarding results to aggregators.


\subsection{Motivation and Contributions}
The mentioned studies showcased the effectiveness of partitioning and aggregation techniques for distributing DNN models. Nevertheless, we still need to investigate how models can collaborate, exchange knowledge, and utilize each other's computational resources.

In this paper, we propose a coordinated DNN algorithm at the edge, namely CoDE, to enhance the services offered by SPs. CoDE consolidates individual DNN models on a unified platform by knowledge transferring among them \cite{transfer-survey} and enhances DNN services by employing resources from other DNN services while ensuring the integrity of DNN models.
 
Our proposed algorithm establishes new computational and inference paths by linking different DNN models. When SPs are under heavy load, they can use the paths to reduce their computation by offloading the tasks. In this way, SPs increase the system capacity and maintain their quality of services (QoS) \cite{edge-qos}. In our algorithm, DNN services can bypass their own obstacles using these pathways, as illustrated in Fig. 2b, while coordinating with other SPs.

Sharing the models may compromise  the privacy of SPs. To address this concern, our proposed algorithm does not require the SPs to publish their model’s types or parameters. For example, host SPs may indicate support for images or text without revealing task specifics or disclosing model architecture (e.g., CNN, Transformer) without exposing parameters.

Our main contributions are as follows:

\begin{itemize}

\item We introduce coordination among DNN models within or across SPs to allow task offloading to other DNN models and provide more service options. We freeze the models' parameters and split them by a number of blocks. We add learnable links among the models' layers to create new paths for the tasks. Our algorithm reduces redundant parameters by preventing replication and avoiding EEs.
 
\item We enable the DNN services to either skip their blocks directly by using skip-connections, or use other services' blocks by using cross-connections. These connections help DNN services decrease their local computation.
 
\item We then propose CoDE that facilitates the coordination of DNN model sharing and SPs' resource utilization. Our algorithm obtains the optimal paths when maximizing a reward function which reflects the model accuracy and the inference throughput. The algorithm compares the reward with that of the original model so as to either add the new path to the system or discard it.

\item We conduct four experiments to assess CoDE's performance. Results show CoDE can generate paths with a slight decrease in precision with no extra local parameters. In particular, CoDE increases the local service throughput by up to $40\%$ while the average accuracy is reduced by only $2.3\%$. Compared to the EE method \cite{edge-early}, it achieves superior accuracy with less local computation.



\end{itemize}

This paper is organized as follows. In Section II,  we introduce the system model and the proposed algorithm. In Section III, we explain the experiments and provide the results. Section IV concludes the paper.

\section{Coordinated DNN on Edge (CoDE)}

In this section, we first present the system model and linking blocks. We then propose our coordinated DNNs on edge algorithm which we call as CoDE.

\subsection{System Model}
Each SP offers one or more DNN services, and each DNN service comprises a DNN model. We split pre-trained DNN models into $N$ DNN blocks, as shown in Fig. \hyperref[fig:fig2]{2a}. We maintain model integrity by freezing the parameters of these blocks. It means that these parameters are not getting updated. Each block can make connections with either local front blocks or host blocks with links, as can be seen in Fig. \hyperref[fig:fig2]{2b}. Links are small neural network modules with learnable parameters.
Two types of connections are defined as follow:
\begin{itemize}
    \item \textbf{Skip-connection}: It connects local blocks to blocks placed ahead in the path with a link, allowing them to skip over the local blocks in the path.

    \item \textbf{Cross-connection}: It connects two different DNN models with links. These links are established between the host blocks that are included in the generated path. All of the cross-connection links are located on the host server.
\end{itemize}

\begin{figure}[t]
    \label{fig:fig3}
    \centering
    \includegraphics[scale=.9]{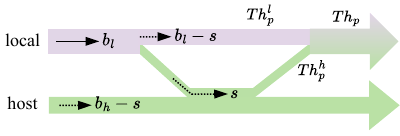}
    \vspace{-.4cm}
    \caption{The host service reserves $s$ samples of its batch. The total throughput is the sum of the local and host throughput (i.e., $Th_p =Th_p^l + Th_p^h$).}
    \label{fig:enter-label}
    \vspace{-.1cm}
\end{figure}

SPs design the links' architecture for their DNNs based on preceding and succeeding block sizes and their architectures. For example, if the SP aims to make skip-connections \cite{switch} for one of its DNNs, it can use any links. Otherwise, if it aims to use the host blocks, it just sends data to the host SP, and it takes care of making the new links between the DNN blocks requested by the local SP.
A cross-connection allows the local service to offload its computation partially on the host service. On the other hand, a skip-connection can be established on a single service locally to improve efficiency.
The batch size of the local and host services are denoted as $b_l$ and $b_h$, respectively. The local service offloads $s$ samples of its batch on the path. To accommodate this offloaded data, the host service reserves space for these $s$ samples when a new cross-connection is established, as shown in Fig. \hyperref[fig:fig3]{3}.

\subsection{Linking Blocks}


The local and host SPs partition their models into $N_l$ and $N_h$ blocks, respectively. Let $\mathcal{P}$ denote the set of all possible paths while $p \in \mathcal{P}$ represents a path. Each path $p \in \mathcal{P}$ is represented by $\mathbf{r}_p = [lout_p, hin_p, hout_p, lin_p]$ where $lout_p$ denotes the local block that connects to the $hin_p$ block in the host app with the corresponding link, as shown in Fig. \hyperref[fig:fig2]{2b}. Similarly, the $hout_p$ block in the host app connects to the $l{in}_p$ block in the local app. We ensure that $0 \leq l{out}_p < l{in}_p < N_l$ and $0 \leq h{in}_p \leq h{out}_p < N_h$. Alternatively, skip-connections only involve $l{out}_p$ and $l{in}_p$ since the local server does not communicate with the host. For such cases, we assign $h{in}_p = h{out}_p = N_f$, where $N_f \gg N_l$.

We define the \textit{inference throughput} as the number of predictions a model can make in a given time. The local service has two processing streams: A main stream for its own workload, and a host stream for the offloaded tasks. It allows the service to process its own tasks faster by reducing computation per batch, resulting in an enhanced throughput. We define the total throughput as $Th_{p}$, which is the sum of the main and host streams throughput as denoted by $Th_{p}^l$ and $Th_{p}^h$, respectively. The block processing time allows SPs to determine the throughput. Also, SPs can simply calculate them with a test batch.
We introduce the average accuracy $A^{av}_p$ and total throughput $Th_p$ for each path $p \in \mathcal{P}$ as follows:
\vspace{-.15cm}
$$Th_p =  Th_p^l + Th_p^h$$
\begin{equation}\label{equ:Ap}
    A^{av}_p = (Th_p^l A_0 + Th_p^h A_p) / Th_p,
\end{equation}
where $A_p$ and $A_0$ are the accuracy of path $p$ and the main path, respectively. We define function $\text{F}(p)$ for each $p \in \mathcal{P}$ to represent its reward as below:

\begin{figure}[t]\label{fig:fig4}
\includegraphics[width=.5\textwidth,keepaspectratio]{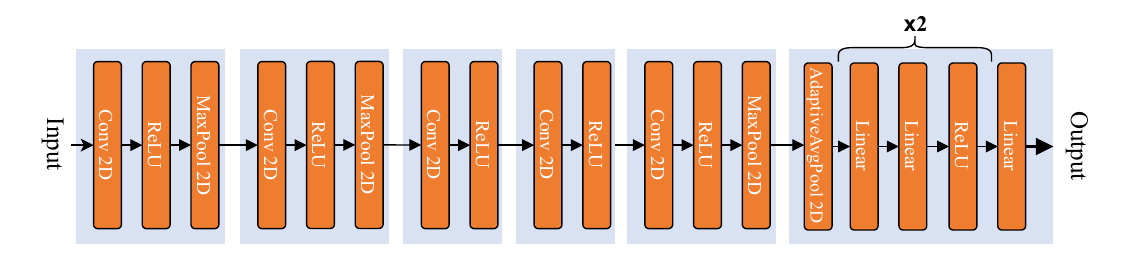}
\vspace{-.6cm}
\caption{A sample of a partitioned AlexNet model with $N=6$.}
\end{figure}

\vspace{-.2cm}
\begin{equation}
    \text{F}(p) = \xi(A^{av}_p) \zeta(Th_p),
\end{equation}
where $\xi(A^{av}_p)$ is defined as the accuracy reward function and $\zeta(Th_p)$ is the throughput reward function. In order to reward the desired accuracy range $A^{av}_p>A_{min}$, where $A_{min}$ is set by the local SP, we choose $\xi(A^{av}_p) = Sigmoid(k(A^{av}_p - A_{min}))$, where $k$ is a constant. We also define the throughput reward function $\zeta(Th_p) = Th_p - Th_0$ that shows the additional throughput path $p$ can provide compared to the original throughput $Th_0$.
The number of the possible paths is $\mathcal{O}(N^4)$. Searching through the entire space would take a long time since we need to train all the paths. Hence, we employ a mechanism to predict the accuracy values $A_p$. We use a multi-stage optimization algorithm where, at each stage $n$, we aim to estimate the optimal path, denoted as $p^n$, using the $n-1$ previously calculated accuracy values of the paths as denoted by set $\mathcal{P}^{n-1}=\{p^1,..., p^{n-1}\}$.
    
We define the distance between $p^i$ and $p^j$ as  $d_{p^i, p^j}=\|\mathbf{r}_{p^i} - \mathbf{r}_{p^j} \|_2$.  By subtracting $\mathbf{r}_{p^j}$ from $\mathbf{r}_{p^i}$, we can identify the blocks that exist in only one of the paths $p^i$ and $p^j$. We predict the accuracy of path $p^n$ which is shown by $A_{p^n}$ with a weighted average of paths in $\mathcal{P}^{n-1}$. The predicted value is:

\vspace{-.35cm}

\begin{equation}
\begin{aligned}
     \hat{A}_p = Pred_n(A_p) = \frac{\sum_{i=1}^{n-1} w^i_p A_{p^i}}{\sum_{i=1}^{n-1} w^i_p}, & ~~ w^i_p = 
     \begin{cases}
    \frac{1}{d_{p, p^i}},& p\neq p^i\\
    0,              & \text{o.w.}
\end{cases}
\end{aligned}
\end{equation}

The weighted approach leverages the correlation between paths, as closer paths often lead to similar outcomes. Additionally, we calculate $\hat{A}_p^{av}$ similar to (\ref{equ:Ap}).
Thus, we formulate the following problem for stage $n$ which aims to maximize the reward function:
\begin{equation}
\begin{aligned}
     \underset{p}{\text{maximize }} & \xi(\hat{A}^{av}_p) \zeta(Th_p) \\
     \text{subject to } & Th_p > Th_0, \\ & \hat{A}^{av}_p= (Th_p^l A_0 + Th_p^h A_p) / Th_p, \\ & p \in \mathcal{P}.
\end{aligned}
\end{equation}

\renewcommand{\algorithmicrequire}{\textbf{Input:}}
\renewcommand{\algorithmicensure}{\textbf{Output:}}
\newcommand{\myEndIf}{\textbf{end if}}
\algtext*{EndIf}{} 
$~$
\vspace{-.4cm}
\begin{algorithm}[h]
\caption{CoDE}\label{alg:cap}
\begin{algorithmic}[1]
\Require $k, N_l, N_h, b_l, b_h, s, A_{min}, c_{stop}, \epsilon$
\Ensure $p^*$
\State Calculate throughput for all $p \in \mathcal{P}$
\State $\mathcal{P}' = \{ p | p \in \mathcal{P}, Th_p > Th_0 \}$
\State $q \gets 0 ,n \gets 0 , c \gets 0, q_{prv} \gets -1, \quad Q \gets \emptyset$
\While{$c \neq c_{stop}$}
    \State $n \gets n + 1$
    \For{ $p \in \mathcal{P}'$}
    \State $\hat{{A}} \gets Pred_n(A_p)$ \Comment{Predict accuracy} \vspace{.1cm}
    \State $\hat{{A}}^{av} \gets (Th_p^l A_0 + Th_p^h \hat{A}_p) / Th_p $ 
    \State $Q \gets Q \cup \xi(\hat{A}^{av}_p) \zeta(Th_p) $
    \EndFor
    \State $p^* \gets argmax(Q)$
    \State $A_{p^*} \gets$ Accuracy of the trained $p^*$
    \State $q \gets F(p^*)$
    \If{$q - q_{prv} < \epsilon$} 
    {$c\gets c+1$\;} \textbf{else} { $ c \gets 0$\;} \textbf{end if} \EndIf
    \State $q_{prv} \gets q$
    \State $\mathcal{P}^{n} \gets \mathcal{P}^{n-1} \cup p^*$
\EndWhile
\end{algorithmic}
\end{algorithm}

\vspace{-.2cm}
The above problem is a combinatorial optimization problem. As the number of possible paths increases, the complexity of solving this problem escalates tremendously. The complexity of solving such problem would be $O(2^{|\mathcal{P}|})$ in general.
Thus, we propose CoDE, as presented in Algorithm \ref{alg:cap}, to overcome the complexity of this problem. At each stage, the algorithm predicts $F(p)$ for all paths with a total throughput higher than $Th_0$. Subsequently, it predicts the path with the highest reward, trains that path, and calculates its actual reward. If the algorithm fails to achieve higher rewards after $c_{stop}$ stages, it terminates the search. The desired path will be generated by linking the blocks of the local and host DNN services.
This algorithm handles both cross- and skip-connections. However, by adding a skip-connection, we increase the local service's parameter count.

By using Algorithm \ref{alg:cap}, all the links of the cross-connections are positioned on the host DNN services, and the backpropagation process terminates after the entry link. This approach allows services to expand their functionalities without additional parameters on the local device. Generally speaking, it fosters a more flexible network with enhanced capabilities through collaboration among DNN services.

\section{Experiments}

We conduct four experiments to evaluate the performance of our proposed algorithm. Through these experiments, we verify that our approach greatly enhances the overall performance via selective offloading and shortcut routes at the cost of a slightly reduced accuracy, when compared to EE method \cite{edge-early}.

\subsection{Experiment 1: AlexNet - AlexNet} \label{exp1}

\begin{figure}[t]
\centering
\includegraphics[width=.5\textwidth,keepaspectratio]{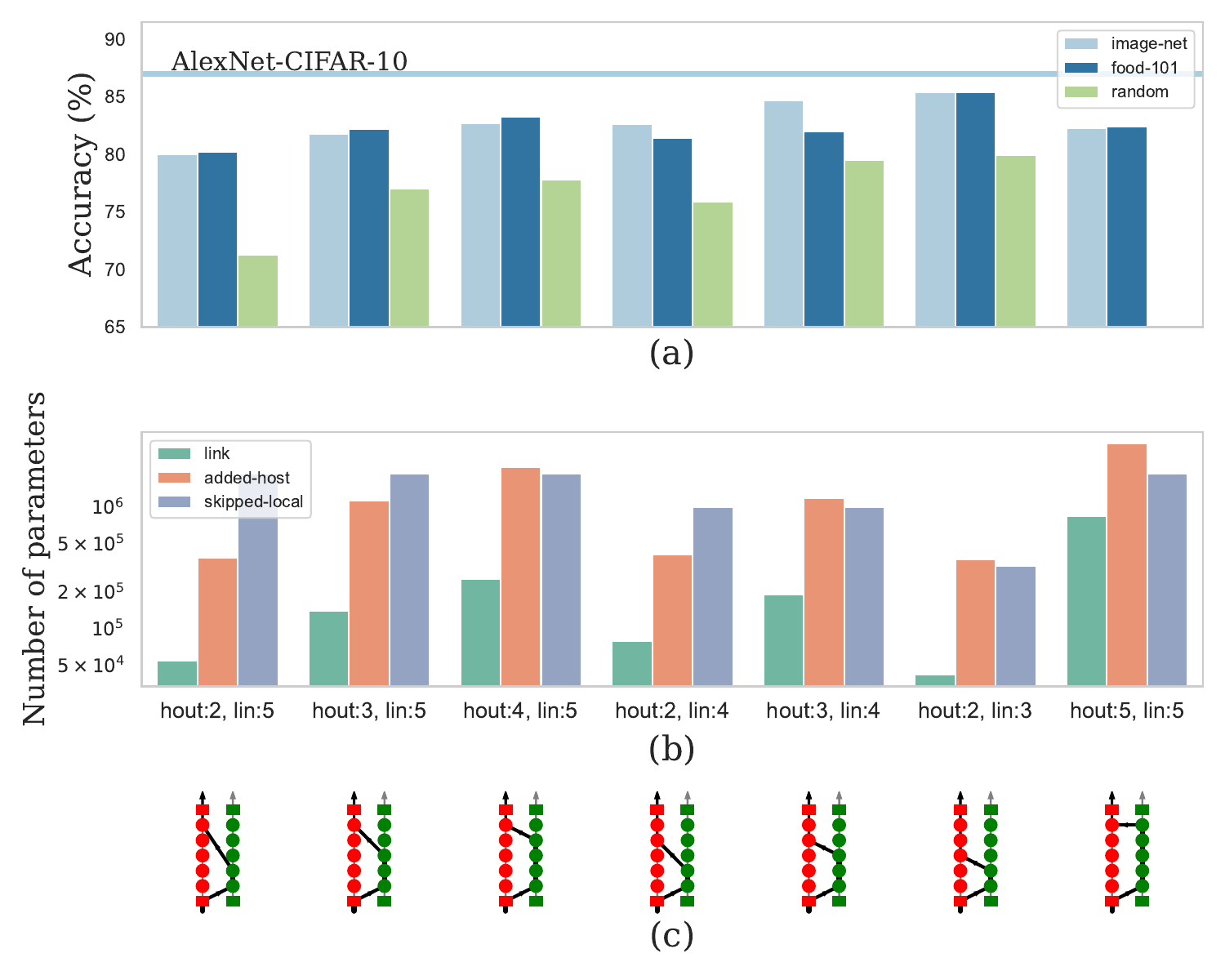}
\label{fig:fig5}
\vspace{-.9cm}
\caption{(a) The accuracy of generated paths. We set $lout_p=0$ and $hin_p=1$. The local model is AlexNet, and it is optimized for CIFAR-10, and its accuracy is $86.7\%$. The host model is also AlexNet, which is optimized for Image-net and Food-101. We also measure the performance of a random model to compare with pre-trained models. (b) The number of parameters when we add a new path. The number of parameters associated with the links is relatively low, but the number of the local-skipped and host-added parameters (sum of the links' and host's parameters) are higher and change according to the paths. (c) These DNN models show the related structure for each connection.}
\end{figure}

In this experiment, both models are AlexNet \cite{alexnet}, with a cross-connection established between them. The objective is to evaluate cross-connection performance for two DNN services with the same architectures but different tasks, depicted in Figs. \hyperref[fig:fig5]{5a} and \hyperref[fig:fig5]{5b}. Fig. \hyperref[fig:fig4]{4} shows a 6-block architecture, but we add another block at the beginning of the model. The local model is optimized for CIFAR-10, and its accuracy is 86.7\%. On the other hand, the host model is optimized separately for ImageNet and Food-101 datasets to assess the performance across different tasks. In addition, to measure the effect of the models, we use random parameters in another task with the same setup. We set the $lout_p$ and $hin_p$ to 0 and 1, respectively, and conduct the experiments when varying the $l{in}_p$ and $hout_p$, which means that we skip blocks 1 to $lin_p$ in the local app. We also vary the number of host blocks from 1 to $hout_p$. All the links are 2D CNN layers. When the actual output size of a block is not equal to the input size of the next block of the path, the transition between the local and the host app or skip-connections, we use a 2D CNN layer.

Fig. \hyperref[fig:fig5]{5} shows the accuracy, number of parameters, and the model architectures. The results show a considerable difference between the pre-trained host models and the random model. With $hout_p=2$ and $l{in}_p=5$, we reduce $1.88$ million local parameters, and we achieve an accuracy of $80.2\%$ with Food-101 and $80\%$ with ImageNet. However, the accuracy for the random host model is $71.3\%$. It means that if considering the random model as the baseline for the accuracy drop, we can compensate $58\%$ for the accuracy drop with the pre-trained models. Furthermore, if we consider the skip-connection from \textit{lout:0} to \textit{lin:5} with accuracy of $61\%$ as the baseline, then we can compensate $78\%$, which is significant.

If we increase the host blocks, we do not necessarily raise the paths' precision. For example, when $hout_p$ is raised from 4 to 5, the accuracy drops from $83.3\%$ to $82.4\%$, contradicting assumptions in EE models. In \cite{edge-early}, the author used early exit on VGG-16, which performs more powerfully than AlexNet on CIFAR-10, but its accuracy is dropped from $93.5\%$ down to $80\%$, and even EEs add a large number of parameters in the local servers. Our method achieves improvements in both local computation and model accuracy compared to \cite{edge-early}.

\begin{figure}[t]
\includegraphics[width=.5\textwidth,keepaspectratio]{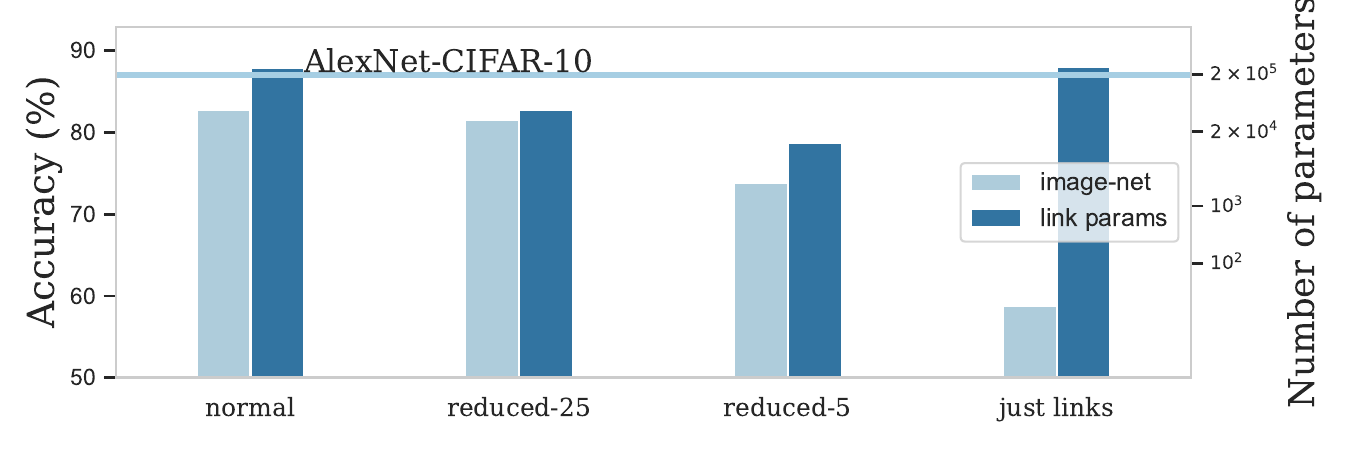}
\label{fig:fig6}
\vspace{-.9cm}
\caption{Accuracy for different links' parameters. Reduced-25 and reduced-5 shrink the link's parameters by $5$ and $25$ times, respectively. Just-links approach measures the impact of the links by removing the host blocks.}
\end{figure}

We also assess the effect of the link size on the path performance by reducing CNN filter numbers using two CNN layers. The first layer decreases input layers to either 25 or 5, as shown in Fig. \hyperref[fig:fig6]{6}, and then the next CNN layer extends it to the next block's input size. In these cases, if we set $\textbf{r}_p = [0,1,4,5]$, links shrink by almost $5$ and $25$ times, respectively. By setting the layer's filter size to 25, the accuracy decreases from $82.7\%$ to $81.5\%$, and if we set it to 5, then the accuracy would be $73.8\%$. As a result, SPs can adjust their performance, not just by changing the number of blocks but also with the link size.

To investigate the host models' effect on the accuracy, we then remove the host blocks and just use the relevant links to measure the effect of the links in the performance alone. The accuracy is significantly reduced to $58.8\%$ because most of the links are just bottleneck CNN layers, which cannot effectively extract the input's features.

\begin{figure*}[h]
\centering
\includegraphics[width=0.85\textwidth,keepaspectratio]{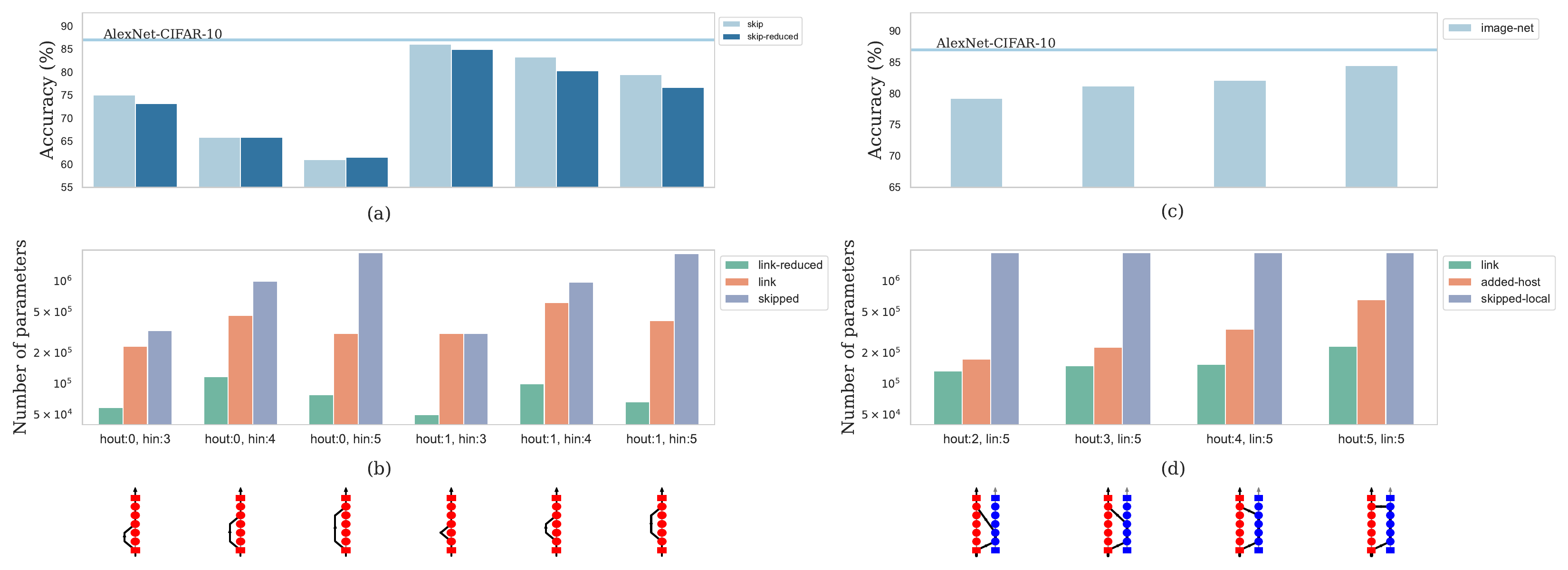}
\label{fig:fig7}
\vspace{-.4cm}
\caption{(a) The accuracy in the skip-connection mode, where the first layers have a greater impact as they calculate the low-level features. (b) The link and link-reduced show the number of the learnable parameters in the links, and the skipped shows the number of the parameters. (c) Here, we change the host model architecture to MobileNet. Again, we can revive a decimated performance by using cross-connection. (d) The number of the parameters of the skipped blocks of the local model and the added parameters from the host model.}
\vspace{-.3cm}
\end{figure*}

\subsection{Experiment 2: AlexNet - Skip-connection}

In this experiment with an AlexNet model, skip-connections are established between its blocks. Additionally, we investigate how the number of link parameters affects path performance. In this experiment, links are either one CNN layer in normal mode or two CNN layers with one MaxPool layer in reduced mode. In the normal mode, links are generated in order to convert the output size of the blocks, but in the reduced mode, we use two CNN layers to reduce the number of filters.

When we set $lout_p$ to 0 for skip-connections, the accuracy notably decreases. It means that the links cannot be learned properly. On the other hand, when we set $lout_p$ to 1, the performance rises, and as the gap between $lout_p$ and $lin_p$ increases, the accuracy drops more.

Fig. \hyperref[fig:fig7]{7a} shows that if we reduce the link size by 5 times, it decreases the accuracy slightly. For example, when we set $lout_p=1$ and $lin_p=5$, the accuracy would be $83.3\%$ and $80.3\%$ for the normal and reduced mode, respectively. Both modes have comparatively fewer parameters than the skipped parameters. Fig. \hyperref[fig:fig7]{7b} shows the number of the link's parameters and the skipped parameters of the main model. All in all, skip-connection performs less effectively in parameter reduction compared to cross-connection. With $lout_p=1$ and $lin_p=4$, inference time is $1.4 ms$, $30\%$ faster than the main model's inference time of $2$ms.


\subsection{Experiment 3: AlexNet - MobileNet}

In this experiment, we use MobileNet \cite{mobile} as the host DNN service. Fig. \hyperref[fig:fig6]{6} shows that we can compensate for the performance degradation with a host model, even with a different architecture. The host model is optimized for ImageNet \footnote{PyTorch pre-trained model}, as shown in Fig. \hyperref[fig:fig7]{7c}. In this experiment, $lout_p$, $lin_p$, and $hin_p$ are set to 0, 5, and 1, respectively, and $hout_p$ is varied from 2 to 5. Here, the accuracy increases monotonically with the rise of $hout$, which did not happen in Experiment 1. In this scenario, again the number of the added-host parameters, which is the summation of the link's and the added host's parameters, is lower than the skipped-local parameters, as shown in Fig. \hyperref[fig:fig7]{7d}.

By using MobileNet for ImageNet, we can revive the accuracy up to $84.5\%$, which is sufficiently high to serve as a coordinated model.  This means, we reduce local computation by $75\%$ just by $2\%$ reduction in the overall accuracy.

\begin{figure}[t]
\includegraphics[width=.5\textwidth,keepaspectratio]{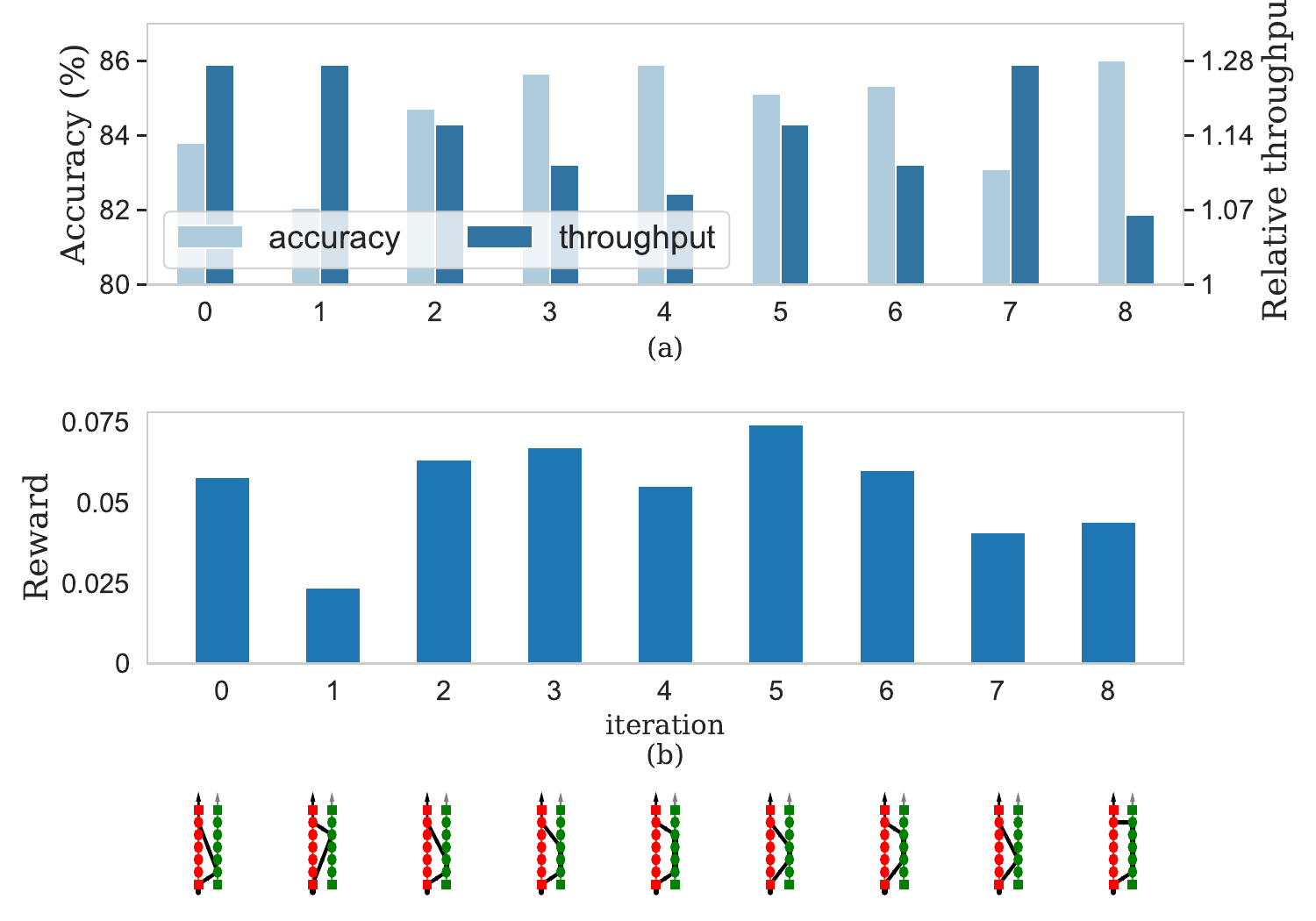}
\label{fig:fig8}
\vspace{-.8cm}
\caption{(a) Accuracy and relative throughput for different cross-connection paths between the local and host models. (b) The reward of the paths according to the proposed setups. In iteration 4 when $lout_p=0$, $hin_p=1$, $hout_p=4$, and $lin_p=5$ has the highest value.}
\vspace{-.3cm}
\end{figure}

\subsection{Experiment 4: Selecting paths}

We now explore the impact of accuracy and computation time on path selection. We use two P100 GPUs on the same machine, where the stream's delay is negligible. In order to calculate the throughput of each path, we use the computation time of each block. We also consider that the host architecture is AlexNet, and use the proposed algorithm and the reward function to find a path. For this test, we set $b_l=b_h=32$, $N_l=N_h=6$, $s=8$, $A_{min}=86\%$, $c_{stop}=3$, and $\epsilon=0.01$.

In this experiment, $\mathbf{r}_p=[0,1,4,5]$ exhibits the best overall performance, as can be seen in Fig. \hyperref[fig:fig8]{8}, with average accuracy $84.4 \%$ and $Th_p/Th_0 = 1.4$. In simpler terms, this means that SP1 can generate a path with a $40\%$ increase in throughput and only a $2.3\%$ decrease in the total accuracy. As a result, it is highly efficient for SP1 to generate a path with a host model optimized for ImageNet.

\section{Conclusion}

In this paper, we studied AIaaS and introduced cooperation among different SPs as well as their DNN models. SPs can use other SPs' resources to offload their computations without replicating their DNN models on them. The local and host SPs do not necessarily need to have the same DNN models and architectures. We first formulated an optimization problem to find the paths that result in the highest amount of reward. To solve this problem, we then proposed a task-offloading approach called CoDE that facilitates collaboration among the SPs and their DNN models. We further conducted four different experiments to investigate the performance of CoDE. The results show that CoDE can significantly increase the inference throughput compared to the EE methods while degrading the accuracy by a small amount. In future, we will investigate how we can manage the traffic flow dynamically through the paths based on request characteristics, such as time sensitivity and inference deadline.

\bibliography{main}
\bibliographystyle{IEEEtran}

\end{document}